\def \JMP {{\it J. Math. Phys. }}

\def \bc {\begin{center}}
\def \ec {\end{center}}

\def \bfr {\begin{flushright}}
\def \efr {\end{flushright}}

\def \caja {\makebox[3.2cm][1]}
\def \v {\vskip}
\def \ii {\'\i}

\def \ba {\begin{array}}
\def \ea {\end{array}}

\def \bea {\begin{eqnarray}}
\def \eea {\end{eqnarray}}

\def \be {\begin{equation}}
\def \ee {\end{equation}}

\def \L {{\cal L}}
\def \R {{\it R}} 
\def \d {\hbox{d}\,}

\def \p {\partial}

\def \w {\omega}
\def \H {{\cal H}}
\def \x {{\bf x}}
\def \q {{\bf q}}

\def \st {\vskip5mm\Large\noindent}
\def \subst {\vskip3mm\large\noindent}

\documentstyle[12pt,a4]{article}

\begin{document}


\thispagestyle{empty}


\noindent hep-th/9503068
\hfil\caja{{\bf Imperial-TP/94-95/25}}\break

\v20mm
\begin{center}

{\bf COMMENTS ON GOOD'S PROPOSAL
FOR NEW RULES OF QUANTIZATION}
\v5mm
 $Miguel$ $Navarro$\footnote[1]{On leave of absence
from {\it Instituto Carlos I de F\'\i sica Te\'orica y Computacional},
Facultad  de  Ciencias, Universidad de Granada, Campus de Fuentenueva,
18002, Granada; and
{\it IFIC, Centro Mixto Universidad de
Valencia-CSIC}, Burjassot 46100-Valencia. SPAIN.}

\v5mm

 The Blackett Laboratory, Imperial College, Prince
Consort Road, London SW7 2BZ; United Kingdom.

\end{center}

\v5mm
\centerline{\bf Abstract}
\v3mm
\footnotesize

In a recent paper \cite{[Good1]}
Good postulated new rules of quantization, one of the major
features of which is
that the quantum evolution of the wave function is always
given by ordinary differential equations.
In this paper we analyse the proposal in some detail and discuss
its viability and its relationship with the
standard quantum theory.
As a byproduct, a simple derivation of the `mass spectrum'
for the Klein-Gordon field is presented, but it is also shown that there
is a complete additional spectrum of negative `masses'.
Finally, two major reasons are presented against the viability of
this alternative proposal:

a) It does not lead to the correct energy spectrum for the hydrogen atom.

b) For field models, the standard quantum theory cannot be recovered
 from this alternative description.
\v20mm
\noindent PACS numbers: 03.65.Bz, \ 11.10.Ef, \  03.70.+k

\normalsize
\vfil\eject

\setcounter{page}{1}

\st{\bf I. Introduction}
\v3mm \normalsize
In a recent paper \cite{[Good1]} (see also \cite{[Good2]})
Good has proposed
new rules for (second) quantizing field models.
The necessary postulates of interpretation for this
new formalism are missing in the
paper of Good,  but natural ones can be found  (see below).
The quantization rules proposed
by Good, together with these natural rules of interpretation,
constitute a quantum  theory,
different from the usual or standard one, which we will refer to as `new' or
`finite' quantum theory.

The `new' quantum theory proposed by Good
possesses a number of remarkable properties which
deserve to pay attention to it:

- The rules of quantization treat time and spatial coordinates on the
same footing and are explicitly covariant, giving rise
to a quantum theory that is always explicitly Poincar\'e invariant.
This therefore avoids the proofs of consistency which
are necessary in the standard approach.

- Everything in the quantum theory - equations of motions,
integration measures, etc. - is based on finite-dimensional calculus,
thereby avoiding the problems of infinities
that appear in the standard approach.

Moreover, this formalism would lead to a discrete
`spectrum' of allowed masses for the particles of the quantum theory.

The purpose of this paper is to study what lies
`behind' all these interesting properties and to check whether or not
this alternative proposal has any chance of competing with the
standard one.

Our conclusion is that
in spite of all the advantages of this proposal, it
does not reproduce the experimental data and therefore is
{\it not} an acceptable alternative to
the standard one.

 The paper is organized as follows:
 In Section 2 we analyse the Good's quantization rules
in some detail and put them in the context of the
finite-dimensional covariant approach to field theory.
In Section 3, we develop several topics of the new theory for
several elementary systems and check its experimental predictions.
We also re-derive the `mass' spectrum for the Klein-Gordon field.
Section 4 is devoted to a comparative description of the
interpretation rules of this theory in relation to the standard one.

\st{\bf II. The hard way to a finite-dimensional
quantum field theory and Good's proposal}
\v3mm\normalsize
There are bassicaly two different ways of looking at field theory
from the standpoint of classical mechanics, or, in other words,
 there are two
different ways of considering field theory as a generalization of
classical mechanics. These can be summarized as follow:

\[\ba{lcc}{}&\hbox{\bf Classical mechanics.}
&\hbox{\bf Classical field theory.}\\
\hbox{A.\ \ }&
q_n(t):\>\>\hbox{$n$ discrete label.}&\varphi^a(\x)(t):
\>\>\hbox{$\x$ continuous label.}\\
{}&{}&{}\\
\hbox{B.\ \ }&q_n(t):\>\hbox{Space-time}&
\varphi^a(\x,t):\>\hbox{Space-time}\\
{}&(1+0)-\hbox{dimensional.}&(1+n)-\hbox{dimensional.}\ea\]

In the perspective A the solutions
of the classical equations of motion are sections of an infinite-dimensional
fiber bundle with co-ordinates $(\varphi^a(\x),t)$ - and  base
manifold co-ordinatized by $t$-,
whereas in the perspective B the
solutions of the equations of motion are sections of a
finite-dimensional fiber bundle
co-ordinatized by $(\varphi^a,\x,t)$ - and base manifold
co-ordinatized by $(\x,t)$ -.
Although both of these points of view
lead to the same classical field theories,
this is not the case for the quantum ones: these different
interpretations lead to different quantum theories.
The interpretation A leads
to the standard quantum field theory,
whereas the interpretation B would lead
to a {\it different} quantum field theory,
if any were ever constructed.

Quantum Mechanics is described by wave functions $\Psi(\q,t)$,
the interpretation of which is:
\bc
{\it $|\Psi(\q_0,t)|^2$ is the probability of finding the result $\q_0$
if a measure of $\q$ is made at the time $t$.}
\ec
This interpretation of Quantum
Mechanics, together with
the two ways, summarized above,
of considering Field Theory as a generalization of Mechanics,
would result in the two different descriptions of the
quantum theory that follow:

{\bf A1.} The quantum theory is described by wave functions (functionals)
$\Psi(\{\varphi(\x)\},t)$ the interpretation of which is:
\bc
{\it $|\Psi(\{\varphi^a(\x)\}_0,t)|^2$ is the probability of
finding the result $\{\varphi^a(\x)\}_0$ if a measure of the
configuration of the field is made at the time $t$.}
\ec
[This is the standard description of Quantum Field Theory.]
\v3mm
{\bf B1.} The quantum theory is described by wave functions
$\Psi(\varphi^a,\x,t)$ the interpretation of which is:

\bc
{\it $|\Psi(\varphi^a_0,\x,t)|^2$ is the probability that a measure of
the field $\varphi$ at the point $(\x,t)$ of the space-time gives the
result $\varphi^a_0$.} \ec

There is no {\it a priory} reason for the correct description of the
quantum theory to be the one in A1 and not that in B1.  In fact,
the description in B1 seems to be better than that in A1 because

- It treat space co-ordinates and time  on the same footing
thus providing a more suitable framework to construct, throught
covariant rules of quantization, a covariant quantum theory.

- The wave functions of the theory are proper functions, not
functionals,  avoiding,  from the very beginning,
the problems of infinities of the standard picture.

- The basic questions which it would be suitable to answer are
of a more local nature than those of the standard approach.

However, a (successful) quantum theory based on the
point of view A is known (the standard one) whereas all attempts to
construct a quantum theory making use of the point of view in B have
failed (Recent discussions on this
subject can be found, for instance, in
Ref. \cite{[3]} and references therein).
In fact it is easy to show (see below, Section 4) that the descriptions
B1 and A1 are {\it not} equivalent to each other
and, therefore, if the description in A1 is considered as the right one,
then the description in B1 cannot be correct as well.

Apart from the problems of interpretation, there are additional obstacles
in the construction of a formalism for this
alternative quantum theory.
The main problem in this approach appears to be that there is no natural
notion of Poisson bracket. Hence, in this
covariant formalism, there is no natural way of
obtaining quantization rules with which to
construct the quantum theory from the classical one.
There is a covariant generalization of the
Legendre transform of Mechanics as
well as a covariant Hamiltonian, but the covariant
Hamiltonian equations of
motion cannot be obtained by means of a Poisson bracket.

Given a Lagrangian $\L = \L(\phi^a,\p_\mu\phi^a)$,
the covariant Hamiltonian
$\H$ is obtained by means of the covariant legendre transform:

\be \H = \pi_a^\mu\p_\mu\phi^a -\L\label{a1b}\ee
where the covariant momenta $\pi^\mu_a$ are defined by:

\be \pi^\mu_a= \frac{\p\>\L}{\p\>(\p_\mu\phi^a)}\label{a2}\ee

If we now write the Lagrangian in the
following covariant Hamiltonian form:

\be \L = \pi_a^\mu\p_\mu\phi^a - \H(\phi^a,\pi_a^\mu)\label{a3}\ee
its Lagrange equations of motion will also
have a covariant Hamiltonian form:

\bea \p_\mu\phi^a &=& \frac{\p\>\H}{\p\>\pi^\mu_a}\\
\p_\mu\pi^\mu_a&=& -\frac{\p\>\H}{\p\>\phi^a}\label{a4}\eea
The obstacle now
is that these equations of motion cannot be associated
with a pair \{Poisson bracket, Hamiltonian\}. Therefore, the basic tool
in the standard formalism for
constructing the quantum theory from the classical one is lacking here.

In the light of the above discussion,
let us consider now the proposal of
Good \cite{[Good1],[Good2]}.

The quantization rules are:

I. For a field model defined in a
$(1+n)$-dimensional space-time coordinatized by
$(\x,t)$ and with fields $\varphi^a$, the quantum theory is
described by wave functions $\Psi(\varphi^a,\x, t)$.

[Therefore, the description of the quantum theory is the one in B1.]

II. The correspondence principle is:

\be \pi^\mu_a\Longrightarrow \widehat\pi^\mu_a=
-\hbar^2\frac{\p^2}{\p \varphi^a\p x_\mu}\label{a6}\ee

III. The quantum equation of motion, the analog of the
Schr\"odinger equation, is

\be \H(\phi^a,-\frac{\p^2}{\p \varphi^a\p x_\mu})\Psi(\varphi^a,x^\beta)=
-\hbar^2\frac{\p^2}{\p x^\nu \p x_\nu}\Psi(\varphi^a,x^\beta)\label{a7}\ee

A number of features of these quantization rules
are inmediately noticeable:

- The new quantization rules {\it do not} lead to the usual
quantization rules in the mechanical case - when the `space-time' is
$(1+0)$-dimensional.

Hence, they should not be looked at as a generalization for field theory
of the familiar quantization rules for Mechanics; they are instead
new quantization rules that serve for Mechanics as well as for
Field Theory.
The quantum theory of Mechanics which will be obtained from these new
rules of quantization {\it is not} the usual Quantum Mechanics.
Therefore, since the standard quantization rules have proved extraordinarily
successful in predicting all the known experimental data,
the next step should be to check whether or not
this new quantum theory leads to the same
predictions as the standard one.
We shall return to this point in Section 3.

- The formula (\ref{a6}) and eq. (\ref{a2}) implies the following
equations for the dimensionalities of
the quantities involved [In the sequel we shall make $c=1,\>[c]=0$; and all
dimensionalities will be expressed in terms of $[x]$ and $[m]$.]:

\be [\L]+[x]-[\varphi^a]=[\pi_a^\mu]= 2[\hbar]-[\phi^a] -[x]\Rightarrow
[\L]=2[\hbar]-2[x] \label{a8}\ee

This dimensionality for $\L$ is, in general,
different from that required for the
standard formalism where $[\L]=[\hbar]-(1+n)[x]$. Ref. \cite{[Good1]}
addresses this problem simply by arguing
that the dimensionality of the Lagrangian is relevant only
for the quantum theory - the standard or the new one -
and that the only relevant point here is that the new approach requires
dimensionalities for the physical quantities different from those
in the standard approach. Another solution
for obtaining the correct dimensionalities
 might be to multiply the Lagrangian
for an adequate factor.

Although these positions could be self-consistent in some cases, there is
a different point of view that is equivalent to the second above
and, furthermore, is in greater agreement with
the historical development of (standard) Quantum Mechanics.  The key point
is that the standard quantization rules required the introduction of a
parameter $\hbar$ with dimensionalities $[\hbar]=[m]+[x]$.
Hence,  we should say
that the new rules of quantization also require
the introduction of a new parameter with the adequate dimensionalities.
Moreover, since there are no physical reasons for the right-hand side of the
Schr\"odinger-like equation to be exactly the form above,
we can- and we shall -modify and generalize
the second and  third rules of quantization
by introducing two parameters $\lambda$ and $\beta$ in the following way:

II'. The correspondence principle is:

\be \pi^\mu_a\Longrightarrow \widehat\pi^\mu_a=
-\frac{\hbar^2}{\lambda}\frac{\p^2}{\p \varphi^a\p x_\mu}\>,
\quad[\lambda]=2[m]-[\L]\label{a6b}\ee

III'.  The quantum equation of motion, the analogue of the
Schr\"odinger equation, is

\be \H(\phi^a,-\frac{\hbar^2}{\lambda}
\frac{\p^2}{\p \varphi^a\p x_\mu})\Psi(\varphi^a,x^\beta)=
-\frac{\hbar^2}{\lambda\beta}\frac{\p^2}{\p x^\nu \p x_\nu}
\Psi(\varphi^a,x^\beta)\>,\quad[\beta]=0\label{a7b}\ee

\st{\bf III. The new quantum theory for
the harmonic oscillator, the Klein-Gordon field and the hydrogen atom}
\v3mm\normalsize
In this section, we shall check the experimental predictions of this new
quantum theory. We shall show that it reproduces the
standard energy spectrum for the harmonic oscillator surprisingly well,
but that it does not, however,
reproduce the right spectrum for the hydrogen atom.
The `mass' spectrum for the Klein-Gordon field is also
found in a very simple manner.

\subst{\bf III.I. The harmonic oscillator}
\v2mm\normalsize
The familiar Lagrangian for the harmonic oscillator is:

\be
 \L_{H.O.}=\frac12m\left[\dot \q^2-\w^2 \q^2\right]\>;\quad
\q=(q_1,..,q_D)\label{b1}\ee
It has dimension $[\L]=[m]$.

The Hamiltonian can be written as

\be \H = \frac{{\bf p}^2}{2m} + m\w^2\frac{\q^2}{2}\label{b3}\ee
The Schr\"odinger-like equation takes the form:

\be \left[\left(-\frac{\hbar^2}{2m}\frac{\p^2}{\p \q^2}\right)
\left(-\frac{\hbar^2}{\lambda^2}\frac{\p^2}{\p t^2}\right)
+ m\w^2\frac{\q^2}{2}\right]\Psi(\q,t)
=-\frac{\hbar^2}{\lambda\beta}\frac{\p^2}{\p t^2}\Psi(\q,t)\label{b4}\ee

Stationary wave functions of definite energy $E$ will satisfy:

\be -\hbar^2\frac{\p^2}{\p t^2}\Psi(\q,t)=E^2\Psi(\q,t)\label{b5}\ee
For these functions the (stationary) Schr\"odinger-like equation is:

\be \left[-\frac{\hbar^2}{2 \left\{\frac{m\lambda^2}{E^2}\right\}}
\frac{\p^2}{\p \q^2}
+ \left\{\frac{m\lambda^2}{E^2}\right\}
\frac{E^2}{\lambda^2}\w^2\frac{\q^2}{2}\right]\Psi(\q,t)
=\frac{E^2}{\lambda\beta}\Psi(\q,t)\label{b6}\ee
It is clear from this equation that the allowed energies will be the
solutions of the equation

\be \frac{E^2}{\lambda\beta}
=\hbar\sqrt{\frac{E^2}{\lambda^2}\w^2}\left(n+\frac{D}2\right),
\qquad n\in {\cal N} \label{b7}\ee

The energy levels of the new harmonic oscillator are therefore:

\be E=\pm \beta\hbar\w\left(n+\frac{D}2\right),
\qquad n\in {\cal N}\label{b8}\ee
Except for the duplication in positive and
 negative energies, making $\beta=1$, eq. (\ref{b8}) exactly
reproduces the familiar energy levels of the standard approach.

\subst{\bf III.II. The mass spectrum of the Klein-Gordon field}
\v2mm\normalsize
The discussion in the preceding subsection makes
it very simple to get the mass spectrum for the Klein-Gordon field
already found,  albeit in a rather
cumbersome manner, in Ref. \cite{[Good1]}

The Lagrangian for the Klein-Gordon field is:

\bea \L = \frac12\hbar^2\left(\p_\mu\phi_1\p_\mu\phi_1 +
\p_\mu\phi_2\p_\mu\phi_2\right)
- \frac12m_0^2\left(\phi_1^2 +\phi^2_2\right)\>.\label{b9}\eea

The Schr\"odinger-like equation takes the form:

\bea \left[-\frac12\left(\frac{\p^2}{\p \phi_1^2}
+\frac{\p^2}{\p \phi_2^2}\right)
(-\frac{\hbar^2}{\lambda^2}\frac{\p^2}{\p x^\nu \p x_\nu})
+\frac12 m_0^2\left(\phi_1^2 +\phi^2_2\right)\right]
\Psi(\phi_1,\phi_2,x^\mu)&&\nonumber\\
=-\frac{\hbar^2}{\lambda\beta}\frac{\p^2}{\p x^\nu \p x_\nu}
\Psi(\phi_1,\phi_2,x^\mu)&&\label{b10}\eea
For stationary wave functions:

\bea -\hbar^2\frac{\p^2}{\p x^\nu \p x_\nu}
\Psi(\phi_1,\phi_2,x^\mu)= m^2\Psi(\phi_1,\phi_2,x^\mu)\label{b11}\eea
it can be written in the convenient form:

\bea  \left[-\frac{\hbar^2}{2\hbar^2\lambda^2/m^2}
\left(\frac{\p^2}{\p \phi_1^2}
+\frac{\p^2}{\p \phi_2^2}\right) +
\left\{\frac{\hbar^2\lambda^2}{2m^2}\right\}\frac{m_0^2m^2}{\hbar^2\lambda^2}
\left(\phi_1^2 +\phi^2_2\right)\right]
\Psi(\phi_1,\phi_2,x^\mu)&&\nonumber\\
=\frac{m^2}{\lambda\beta}\Psi(\phi_1,\phi_2,x^\mu)&&\label{b11b}\eea

This equation describes a two-dimensional harmonic oscillator with
frequency $\w^2=\frac{m_0^2m^2}{\hbar^2\lambda^2}$. By direct comparison with
eq. (\ref{b7}) we obtain the following equation for the allowed energies
(masses):

\be \frac{m^2}{\lambda\beta}=\hbar\sqrt{\frac{m_0^2m^2}{\hbar^2\lambda^2}}
(n+\frac22),
\quad n\in{\cal N}\label{b12}\ee
with solution:

\be m= \pm \beta m_0(n+1),\quad n\in{\cal N}\label{b13}\ee
Therefore, in addition to the positive `masses' already found in Ref.
\cite{[Good1]}, there is also a complete, similar spectrum of negative
`masses'.

\subst{\bf III.III. The hydrogen atom}
\v2mm\normalsize
The Hamiltonian for the hydrogen atom is:

\be \H = \frac{{\bf p}^2}{2\mu} + V({\bf r})\label{b30}\ee
where $V({\bf r})$ is the Kepler potential and
$\mu$ is the reduced mass of the interacting particles.

It is easy to see that the stationary Schr\"odinger-like equation for
this system is:

\be  \left[\frac{-\hbar^2\nabla^2}{2\mu}\frac{E^2}{\lambda^2} +
V({\bf r})\right]\Psi(\x,t)=\frac{E^2}{\lambda\beta}\Psi(\x,t)\label{b31}\ee

A direct comparison with the familiar expression for the standard
hydrogen atom shows that the allowed energies, $E_n$, in eq.
(\ref{b31}) have to be solutions of the equation:

\be \frac{E_n^2}{\lambda\beta}=
\frac{\lambda^2}{E^2_n}\frac{E_{\tiny ground}}{n^2}
\>\Rightarrow E_n^4=\beta\lambda^3 \frac{E_{\tiny ground}}{n^2}
\label{b32}\ee

It is clear that no choice of the parameters $\lambda,\>\beta$ will
reproduce the correct spectrum $E_n=\frac{E_{\tiny ground}}{n^2}$.
Therefore, the new quantum theory
{\it does not} reproduce the right spectrum for
the hydrogen atom.

\st{\bf IV. More on the rules of interpretation}
\v3mm\normalsize
Even though it has been shown in the previous section that Good's
proposal does not predict the correct experimental data, and is
therefore not a valid quantum theory, it could be argued that {\it other}
rules of quantization might lead to a good quantum theory based on the
finite-dimensional covariant description in B1. In fact it is easy
to show that
the rules of interpretation in B1 are not equivalent to those in A1
and, therefore, they would not give rise to an equivalent theory
regardless of the rules of interpretation they were supplemented with.

For the sake of simplicity we shall
consider a space-time of the form $\{1,...,N\}\times \R$; that
is to say, the space has only a finite- or, at most, countable -number
of points.

In the description of the quantum theory in B1 the wave
functions would be of the form

\be \Psi^B=\Psi^B(\varphi,n,t)\equiv\Psi^B_n(\varphi,t)\label{d1}\ee
Thus, there is a function for each point of the space-time.
The interpretation in B1 requires these wave functions to be square
integrable:

\be \int\d \varphi|\Psi^B_n(\varphi,t)|^2 \in\R\label{d2}\ee

This must be compared with the standard description:

\be \Psi^A=\Psi^A(\varphi_1,...,\varphi_N;t)\>. \label{d3}\ee

The condition of square integrability here is:

\be \int\d\varphi_1,...\d\varphi_N
 \,|\Psi^A(\varphi_1,...\varphi_N;t)|^2 \in\R\label{d4}\ee

Now let us assume that the two functions
$\Psi^A(\varphi_1,...,\varphi_N)$ and $\Psi^B(\varphi,n)$ describe
the same physical situation at $t=t_0$. Let us assume in addition
that it is possible, at a fixed time, to measure the field $\varphi$
at any point of the space without disturbing the measurements on
the neighbouring points- this is always
explicitly or implicitly assumed in the description in A1,
whereas in the description in B1 it has
to be considered as an additional postulate -. Then we would have:

\be |\Psi^A(\varphi_1,...,\varphi_N)|^2=
|\Psi^B_1(\varphi_1)|^2...|\Psi^B_N(\varphi_N)|^2\label{d6}\ee
But it is evident that a general function
$|\Psi^A(\varphi_1,...,\varphi_N)|^2$ cannot always be decomposed
as eq. (\ref{d6}) requires. Also, if two functions
$\Psi^A$ and $\Phi^A$ are decomposable, then the linear
superposition of them, $\Psi^A +\Phi^A$,
 which, in the absence of superselection rules,
is also an admissible function, will
not generally admit such a decomposition.

Hence, there are physical situations which can be described with the
standard quantum field theory but which do not admit a description
within the formalism in B1.

\st{\bf V. Final comments}
\v3mm\normalsize
We have shown that
the theory proposed in Ref. \cite{[Good1]}
is not equivalent to the standard quantum theory either for
mechanical systems- it does not reproduce, for instance,
the right energy spectrum for the hydrogen atom -,
or for field models- the physical
interpretation is not the same -. Therefore, that formalism
cannot describe the right `physical' theory.
Nonetheless, the proposal collects such a number of
attractive properties that it is difficult not to believe that
it must contain something true, perhaps as a limiting case of
the standard theory. Therefore, it would be interesting
to find a physical interpretation for it
as well as for the `mass' spectrum it `predicts'
for several fields.

\v5mm

\noindent{\bf Acknowledgements.} The author is grateful to the Imperial
College for its hospitality. The author would like to thank C. Arvanitis for
very useful comments and suggestions.
This works is partially funded by the
Spanish ``Direcci\'on General de Ciencia y Tecnolog\ii a (DGYCIT)''.

\end{document}